\documentclass[twocolumn,secnumarabic,amssymb, nobibnotes, aps, prd]{revtex4-1}

\usepackage{amsmath}
\usepackage{graphicx}

\begin{document}

\title{Folded potentials in cluster physics - a comparison of Yukawa and Coulomb potentials with Riesz fractional integrals}%

\author{Richard Herrmann}%
\email[email:]{herrmann@gigahedron.de}
\affiliation{GigaHedron, Berliner Ring 80, D-63303 Dreieich, Germany}
\date{\today}%
\begin{abstract}
In cluster physics a single particle potential to determine the microscopic part of the total energy of a collective configuration is necessary to calculate the shell- and pairing effects. In this paper we investigate the properties of the Riesz fractional integrals and compare their properties with the standard Coulomb and Yukawa potentials commonly used. It is demonstrated, that Riesz potentials may serve as a promising extension of standard potentials and may be reckoned as a smooth transition from Coulomb to Yukawa like potentials, depending of the fractional parameter $\alpha$. For the macroscopic part of the total energy the Riesz potentials treat the Coulomb-, symmetry- and pairing-contributions from a generalized point of view, since they turn out to be similar realizations of the same fractional integral at distinct $\alpha$ values. 
\end{abstract}

\maketitle
\tableofcontents

\section{Introduction}
Convolution integrals of the type 
\begin{eqnarray}
F(x) &=& \int_{-\infty}^\infty d \xi\, f(x-\xi) w(\xi)  \\
      &=& \int_{-\infty}^\infty d \xi\, f(\xi) w(x-\xi) 
\end{eqnarray}
play a significant role in the areas of signal- and image processing or in the solution of differential equations. 

In classical physics the first contact with the 3D-generalization of a convolution integral occurres within the framework of gravitation and electromagnetic theory respectively in terms of a volume integral to determine the potential $V$ of a given charge density distribution $\rho$:
\begin{eqnarray}
V(\vec{x}) &=& \int_{\textrm{R}^3} d^3 \xi \frac{\rho(\vec{\xi})}{|\vec{x}-\vec{\xi}|} 
\end{eqnarray}
where the weight $w$
\begin{eqnarray}
w(|\vec{x}-\vec{\xi}|) &=& \frac{1}{|\vec{x}-\vec{\xi}|} 
\end{eqnarray}
is interpreted as the gravitational or electromagnetic field of a point charge \cite{jac98, kib04}.

In nuclear physics collective phenomena like fission or cluster-radioactivity, where many nucleons are involved, are successfully described introducing  the concept of a collective single-particle potential, based on folded potentials of e.g. Woods-Saxon type. Several weight functions have been investigated in the past. 

In this paper we will demonstrate, that the fractional  Riesz-potential \cite{riesz, pod99, ort06} which extends the weight function introducing the fractional parameter $\alpha$ 
\begin{eqnarray}
w(|\vec{x}-\vec{\xi}|) &=& \frac{1}{|\vec{x}-\vec{\xi}|^{\alpha}} 
\end{eqnarray}
serves as a serious alternative for commonly used Nilsson- \cite{nie, ni2}, Woods-Saxon \cite{eg} and folded Yukawa potentials \cite{bol72, mol81}, modeling the single particle potential widely applied in nuclear physics as well as in electronic cluster physics.

Hence we give a direct physical  interpretation of a multi-dimensional fractional integral within the framework of fragmentation theory \cite{mar80}, which is the fundamental tool to describe the dynamic development of clusters in nuclear and atomic physics.
\index{folded Yukawa}
\index{Yukawa single particle potential}
\index{Woods-Saxon potential}
\index{fragmentation theory}

\section{Folded potentials in  fragmentation theory }
The use of collective models for a description of collective aspects of nuclear motion 
has proven considerably successful during the past decades. 

Calculating life-times of heavy nuclei \cite{my, gru69, sta12},
fission yields \cite{lus80},  giving insight into phenomena like cluster-radioactivity  \cite{poe102}, bimodal fission \cite{her88} or modeling the ground state properties of triaxial nuclei \cite{t3} -
remarkable results have been achieved by introducing an appropriate
set of collective coordinates, like length, deformation, neck or
mass-asymmetry \cite{mah} for a given nuclear shape and investigating its dynamic properties.
\index{fragmentation theory}
\begin{figure}[!b]
\begin{center}
\includegraphics[width=80mm]{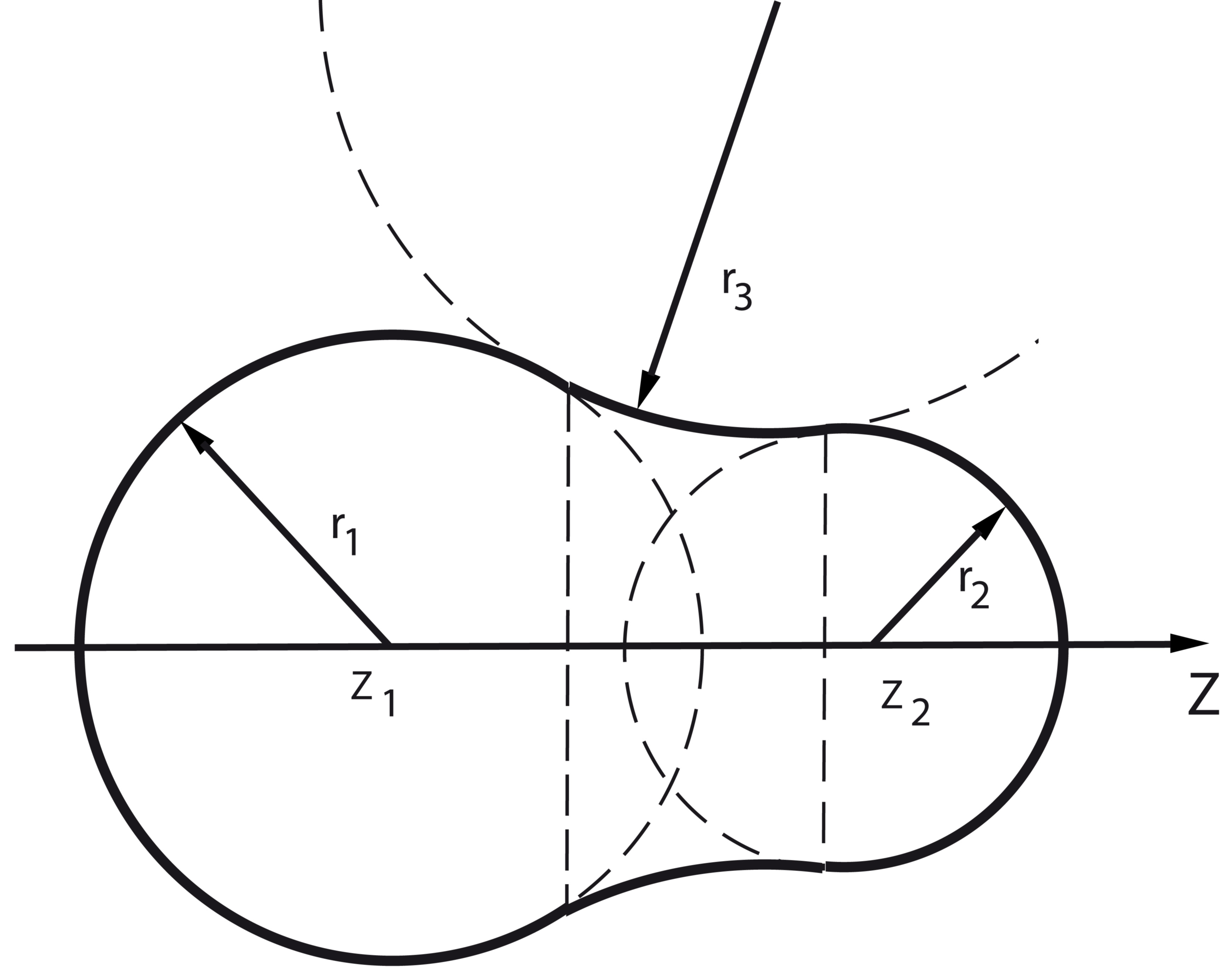}\\
\caption{
\label{fig101}
The parametrization of the 3-spheres-model.  }
\end{center}
\end{figure}

As an example, in figure \ref{fig101} the parametrization of the 3-spheres-model is sketched. It determines the geometry of a given cluster shape by 2 intersecting spheres, which are smoothly connected via a third sphere, which models a neck depending on the size of the radius $r_3$.

\index{collective coordinates}
\index{3-spheres model}
The corresponding set of collective coordinates $\{q^i, i=1,...,4\}$ is given by \cite{dep90}: 
\begin{description}
\item[the two center distance]
\begin{equation}
\Delta z = z_2 - z_1
\end{equation}
\item[the mass asymmetry]
\begin{equation}
\eta_A = \frac{A_1-A_2}{A_1+A_2}
\end{equation}
\item[the charge asymmetry]
\begin{equation}
\eta_Z = \frac{Z_1-Z_2}{Z_1+Z_2}
\end{equation}
\item[the neck]
\begin{equation}
c_3 = \frac{1}{r_3}
\end{equation}
\end{description}
where $A_1, Z_1$ and $A_2, Z_2$ are the number of nucleons  and protons in the two daughter nuclei. 
\index{two center distance}
\index{mass asymmetry}
\index{charge asymmetry}
\index{neck}

This choice of collective coordinates  allows to describe a wide range of nuclear shapes involved in collective phenomena from a generalized point of view \cite{her85}, e.g. a simultaneous description is made possible of general fission properties  and the cluster-radioactive decay of radium
\index{cluster radioactivity}
\begin{equation}
{^{223}}\textrm{Ra} \rightarrow {^{14}}\textrm{C} + {^{209}}\textrm{Pb}
\end{equation}
which was predicted by Sandulescu, Poe\-naru and Grei\-ner in 1980  and later experimentally verified by Rose and Jones in 1984 \cite{san80, ros84}.

In order to describe the properties and dynamics of such a process, 
we start with the classical Hamiltonian function
\begin{equation}
\label{h0h0}
H = T+V_0
\end{equation} 
introducing a collective potential $V_0$, depending on the collective coordinates,
\begin{equation}
V_0(q^i) = E_{\textrm{macro}}(q^i) + E_{\textrm{mic}}(q^i)
\end{equation} 
with a macroscopic contribution $E_{\textrm{macro}}$ based on e.g. the liquid drop model and a microscopic contribution $E_{\textrm{mic}}$, which mainly contains the shell and pairing energy based on a single particle potential  $V_{s.p.}$ and the classical kinetic energy $T$
\begin{equation}
T = \frac{1}{2} B_{ij}(q^i) \, \dot{q}^i \dot{q}^j
\end{equation} 
with collective mass parameters $B_{ij}$.

There are several common methods 
to generate the collective mass parameters $B_{ij}$, e.g.  the cranking model \cite{ing54} or irrotational flow models are used \cite{kel64}.  

Quantization of the classical Hamiltonian\cite{pod28} 
results in the collective Schr\"odinger equation 
\begin{eqnarray}
       \hat{S}_0 \Psi(q^i,t) &=&\Big(-\frac{\hbar^{2}}{2} \frac{1}{\sqrt{B}} \partial_{i}
        B^{ij} \sqrt{B} \partial_{j}  \nonumber \\
&& -i\hbar  \partial_{t} + V_0\Big)\Psi(q^i,t)  = 0
\end{eqnarray}
with $B= \det B_{ij}$ is the determinant of the mass tensor. This is the central starting point for a discussion of nuclear collective phenomena.

For a specific realization of the single particle potential  $V_{s.p.}$, for protons and neutrons respectively a Woods-Saxon type potential may be used. The advantages of such a potential are a finite potential depth and a given surface thickness. Furthermore arbitrary geometric shapes may be treated similarly by a folding procedure, which yields smooth potential values for such shapes. 

For the three-spheres model, in order to define a corresponding potential, a Yukawa-function is folded with a given volume, which is uniquely determined within the model:
\begin{equation}
V_Y(\vec{r}) = -\frac{V_0}{4 \pi a^3} \int_V d^3r^{'} \frac{\exp^{-|\vec{r}-\vec{r}^{'}|/a}}{|\vec{r}-\vec{r}^{'}|/a}
\end{equation}
with the parameters potential depth $V_0$ and surface thickness $a$.
\index{Yukawa function}

For protons, in addition the Coulomb-potential has to be considered, which is given for a constant density $\rho_0$ 
\begin{equation}
V_C(\vec{r}) = \frac{\rho_0}{a} \int_V d^3r^{'} \frac{1}{|\vec{r}-\vec{r}^{'}|/a}
\end{equation}
where the charge density is given by
\begin{equation}
\rho_0 = \frac{Z e}{\frac{4}{3}\pi R_0^3}
\end{equation}
Both potentials may be written as general convolutions in $R^3$ of type:
\begin{equation}
\label{convolAll}
 V_{\textrm{type}}(\vec{r}) =  C_{\textrm{type}} \int_{R^3}  d^3r^{'} \rho(\vec{r}^{'}) w_{\textrm{type}}(|\vec{r}-\vec{r}^{'}|)
\end{equation}
 with the weights
\begin{eqnarray}
      w_{Y}(d)&=& \frac{\exp^{-d/a}}{d/a}\\
      w_{C}(d)&=& \frac{1}{d/a}
\end{eqnarray}
\index{Coulomb potential}
where $d=|\vec{x}-\vec{\xi}|$ is a measure of distance on $R^3$ and a density, which is constant inside the nucleus 
\begin{equation}
 \rho(\vec{r}) = 
\begin{cases}
\rho_0         &  \text{$\vec{r}$ inside the nucleus}\cr
   0 &   \text{$\vec{r}$ outside the nucleus}
\end{cases}
\end{equation} 

Therefore the single particle potential $V_{s.p.}$ is given by
\begin{equation}
      V_{s.p.} = V_Y + (\frac{1}{2} +t_3) V_C + \kappa \vec{\sigma}(\nabla V_Y \times \vec{p})
\end{equation}
with $t_3$ is the eigenvalue of the isospin operator with $+\frac{1}{2}$ for protons and  $-\frac{1}{2}$ for neutrons, which guarantees that the Coulomb potential $V_C$  only acts on protons. The last term is the spin-orbit term with the Pauli-matrices $\vec{\sigma}$, $\vec{p}$ is the momentum operator and the  strength is parametrized with  $\kappa$.  This term is necessary to split up  the degeneracy of energy levels with different angular momentum and to generate the experimentally observed magic shell closures. 

In the original Nilsson oscillator potential an additional $\vec{l}^2$ term was necessary to lower the higher angular momentum levels in agreement with experiment. For Woods-Saxon type potentials such a term is not necessary. Whether  Riesz potentials are a realistic alternative, will be investigated in the next section. 

The solutions of the single particle Schr\"odinger equation with the potential $ V_{s.p.}$ yield the single particle energy levels, which are used to calculate the microscopic part of the total potential energy and contains two major parts, the shell and pairing corrections.
\begin{equation}
E_{\textrm{mic}}(q^i) = E_{\textrm{shell}}(q^i) +E_{\textrm{pair}}(q^i) 
\end{equation}
\index{shell corrections}
\index{pairing corrections}
\section{The Riesz potential as smooth transition between Coulomb and folded Yukawa potential}
If we reinterpret the  Riesz potential 
\begin{equation}
 V_{RZ}(\vec{r}) = C_{RZ} \int_{R^3}  d^3r^{'} \rho(\vec{r}^{'}) w_{RZ}(|\vec{r}-\vec{r}^{'}|)
\end{equation} 
with the weight 
\begin{eqnarray}
      w_{RZ}(d)&=& \frac{1}{(d/a)^{\alpha}} \qquad 0< \alpha < 3
   \end{eqnarray}
as the 3D-version of  the general Riesz integral  (\cite{riesz}) applied to a scalar function $ \rho(\vec{r}) $,  
we may treat and interpret the Coulomb ($\alpha=1$), Riesz- and Yukawa potentials similarly from a generalized point of view. 

\begin{figure}[!hbtp]
\begin{center}
\includegraphics[width=82mm]{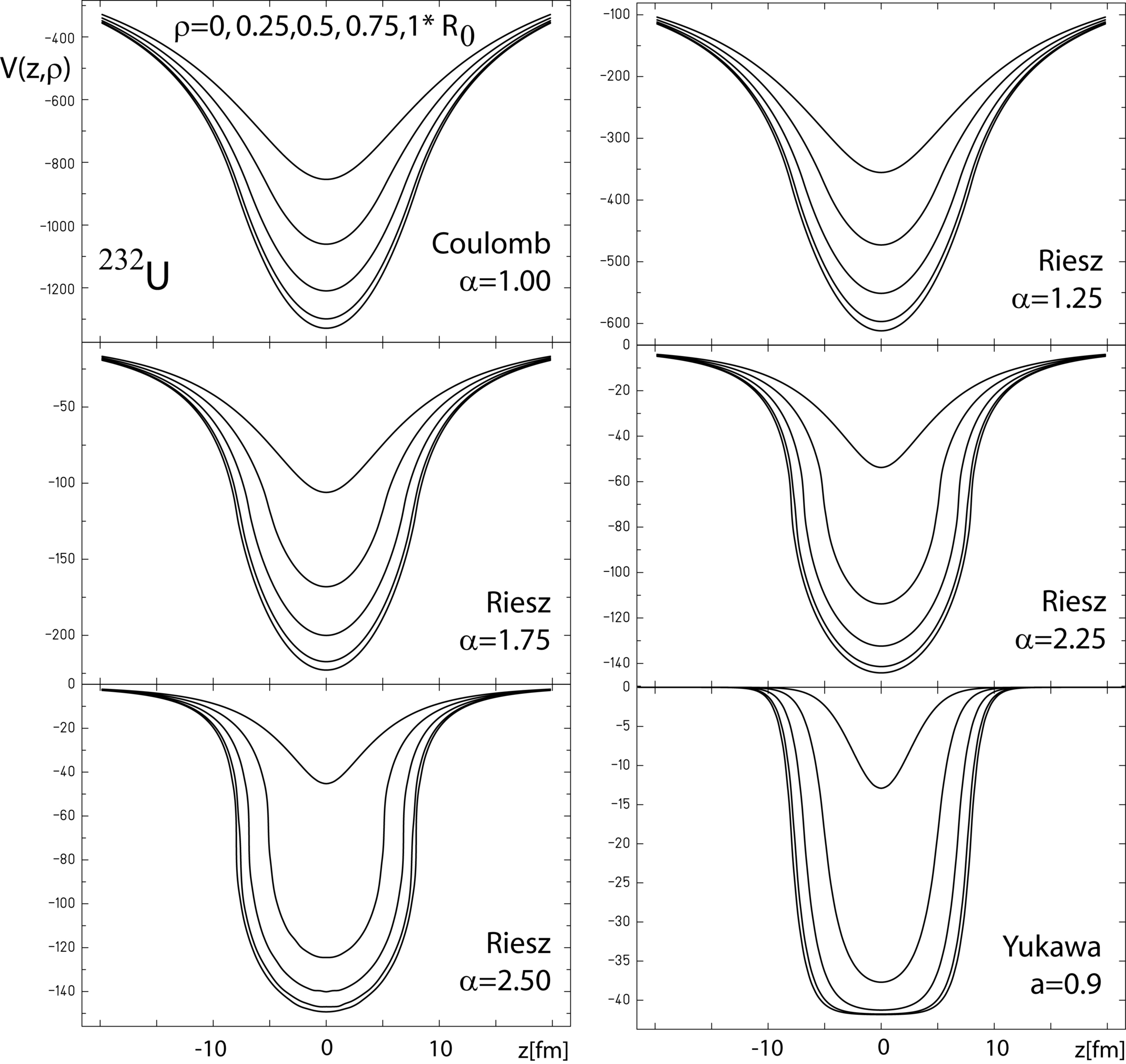}\\
\caption{
\label{potentialRieszsph10}
{
For a spherical assumed shape (here $^{232}$U) the potential for different weight functions is drawn. From top to bottom: Coulomb ($\alpha=1.00$), Riesz ($\alpha=1.25,1.75,2.25,2.50$) and Yukawa (with $a=0.9$[fm] weight. In order to compare the plots with cylinder symmetric shapes,  potential is drawn in cylinder coordinates ($z,\rho$) for a sequence of $\rho=0.00, 0.25,0.50,0.75,1.00 \times R_0$.  $R_0(^{232}\textrm{U}) = 8.26[fm]$  
} }
\end{center}
\end{figure}
\begin{figure}[t]
\begin{center}
\includegraphics[width=82mm]{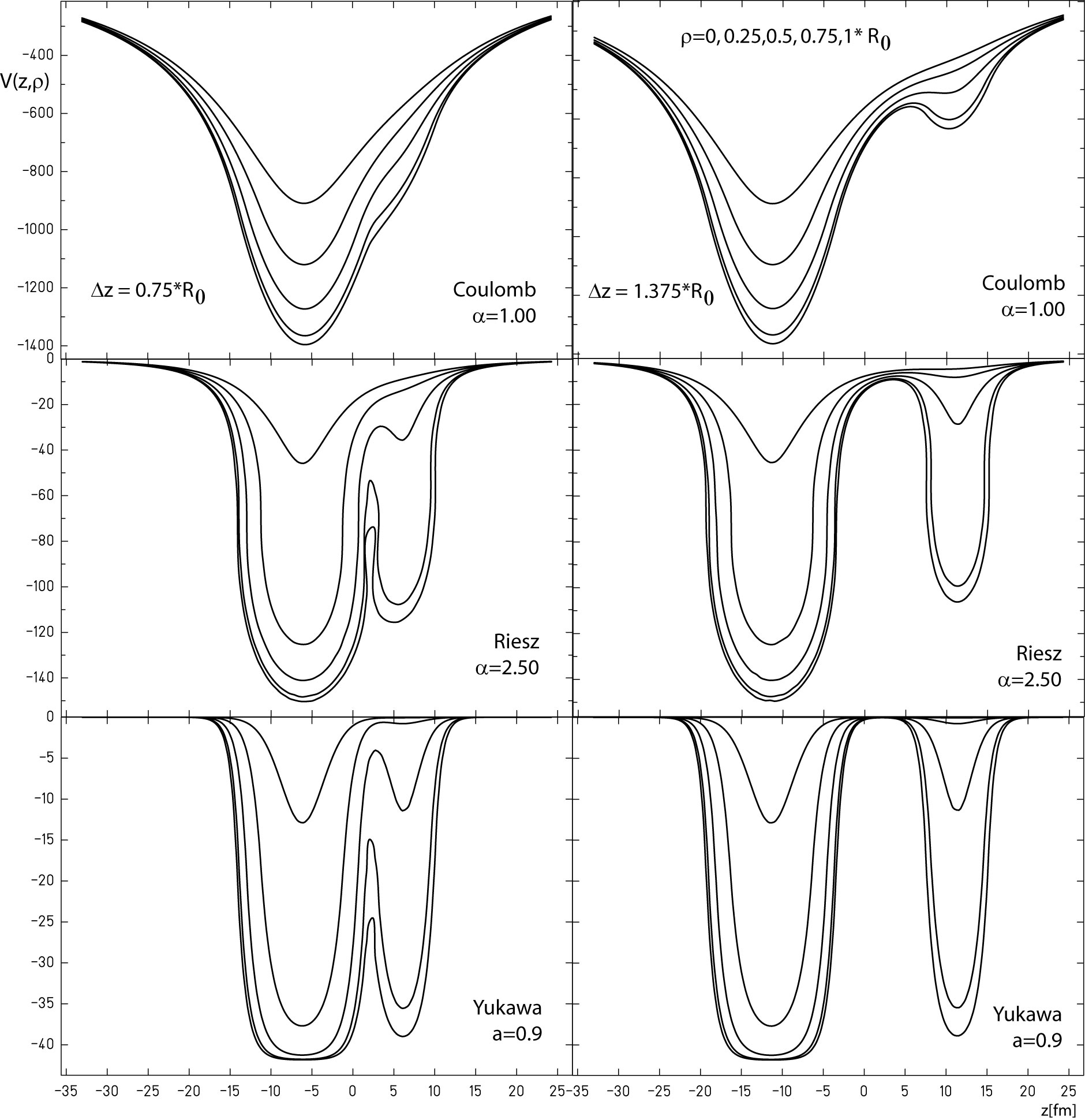}\\
\caption{
\label{Rieszsph3}
{
For the configuration $^{232}\textrm{U} \rightarrow ^{208}\textrm{Pb} + ^{24}\textrm{Ne}$ the Coulomb, Riesz ($\alpha=2.50$) and Yukawa ($a=0.9$[fm]) potential is plotted for  $\Delta z = 0.75 R_0$ (left column) and $\Delta z = 1.375 R_0$ (right column). $R_0(^{232}\textrm{U}) = 8.26[fm]$.
} }
\end{center}
\end{figure}
In the following we will investigate the behavior of the Riesz potential with varying $\alpha$, and compare its properties with the cases of Coulomb and Yukawa weight functions. In a way, the parameter $\alpha$ in the Riesz potential may be interpreted as a global screening of the Coulomb weight, such that the effect of the Yukawa exponential is partly modeled. 
\begin{eqnarray}
\label{conweight1}
      w_{C}(d)&=& \frac{1}{d/a}\\
      w_{RZ}(d)&=& \frac{1}{(d/a)^{\alpha-1}}  \frac{1}{d/a}\\
\label{conweight3}
      w_{Y}(d)&=& \exp^{-d/a} \frac{1}{d/a}
\end{eqnarray}
Therefore the Riesz potential could be an interesting alternative to the Yukawa potential in  the case $\alpha \gg 1 $. In a way, we expect the screening properties of the Riesz potential for increasing $\alpha$ to result in an interpolation between Coulomb and Yukawa limit.

Hence the fragmentation potentials used  in a dynamic description of fission or cluster emission processes are an ideal framework to discuss and understand the properties of the Riesz integral. 

The integral (\ref{convolAll}) with the  weights (\ref{conweight1})-(\ref{conweight3}) may be evaluated analytically for a spherical nucleus with radius $R_0$. 
\begin{equation}
\rho(r) = \rho_0 H(R_0-r)
\end{equation}
with the Heaviside step function $H$.

In this case we have: 
\begin{eqnarray}
 V_{\textrm{type}}^{\textrm{sphere}}(r) &=& C_{\textrm{type}}\, \rho_0 
\int_{0}^{R_0} r^{'2} dr^{'}
\int_{0}^{\pi} \sin(\theta^{'}) d\theta^{'}  \times \nonumber \\
&& \int_{0}^{2 \pi} d\phi^{'}  w_{\textrm{type}}(|\vec{r}-\vec{r}^{'}|)
\end{eqnarray} 
with
\begin{equation}
|\vec{r}-\vec{r}^{'}| = \sqrt{r^2 + {r}^{'2} -2 r {r}^{'}\cos(\theta^{'})}
\end{equation} 
With the substitution $u$
\begin{equation}
u = \sqrt{r^2 + {r}^{'2} -2 r {r}^{'}\cos(\theta^{'})}
\end{equation} 
we end up with a double integral for spherical shapes:
\begin{eqnarray}
 V_{\textrm{type}}^{\textrm{sphere}}(r) &=& 2 \pi  C_{\textrm{type}}\, \rho_0
\int_{0}^{R_0}  dr^{'} r^{'}/r\times \nonumber \\
& & 
\int_{\sqrt{(r-r^{'})^2}}^{\sqrt{(r+r^{'})^2}} du \,u \,w_{\textrm{type}}(u)\\
&=& 2 \pi  \rho_0 \frac{C_{\textrm{type}}}{r} 
\int_{0}^{R_0}  dr^{'} r^{'}\times \nonumber \\
& & 
\int_{|r-r^{'}|}^{r+r^{'}} du\, u\, w_{\textrm{type}}(u)
\end{eqnarray} 
This integral is valid for any analytic weight $w(u)$ and may be easily solved for the Coulomb, Riesz and Yukawa weight functions. We obtain:
\begin{eqnarray}
 V_{\textrm{C}}^{\textrm{sphere}}(r) &=& 
a C_C
\begin{cases}
\frac{Z e}{R_0}(\frac{3}{2}-\frac{r^2}{2R_0^2})         &  \qquad \qquad  \text{$r\leq R_0$}\cr
&\cr
\frac{Z e}{r}                                                             &   \qquad \qquad  \text{$r\geq R_0$}
\end{cases} \\
 V_{\textrm{RZ}}^{\textrm{sphere}}(r) &=& 
4 \pi a^\alpha C_{RZ} \frac{2 \pi}{(\alpha-2)(\alpha-3)(\alpha-4)} \frac{1}{r} \times\nonumber \\
&&
\!\!\!\!\!\!\!\!\!\!
\begin{cases}
 (r + R_0)^{3-\alpha} (r - (3-\alpha) R_0)  +  & \cr
 (R_0-r)^{3-\alpha} (r + (3- \alpha) R_0)                                      &     \text{$r\leq R_0$}\cr
&\cr
(r + R_0)^{3-\alpha} (r - (3 -\alpha) R_0 ) - &\cr
 (r - R_0)^{3-\alpha}  (r +(3 -  \alpha)  R_0)                         &     \text{$r\geq R_0$}
\end{cases} \\
V_{\textrm{Y}}^{\textrm{sphere}}(r) &=& 
4 \pi a^3 C_Y\times\nonumber \\
&&
\!\!\!\!\!\!\!\!\!\!
\begin{cases}
\big(1 - (1+\frac{R_0}{a}) e^{-R_0/a} \frac{\sinh(r/a)}{r/a}\big)         & \!\!\! \text{$r\leq R_0$}\cr
& \cr
\frac{e^{-r/a}}{r/a} \big(\frac{R_0}{a} \cosh(\frac{R_0}{a})-\sinh(\frac{R_0}{a}) \big)    &\!\!\! \text{$r\geq R_0$}
\end{cases} 
\end{eqnarray} 

In figure \ref{potentialRieszsph10} a sequence of these potentials is plotted for a spherical nucleus, ranging from Coulomb ($\alpha=1.00$) and Riesz potential with increasing $\alpha$ up to the Yukawa potential with parameter settings according to \cite{bol72}. 

For large  $\alpha > 2.00$ the Riesz potential as well as the Yukawa potential model a finite surface thickness. 

A remarkable difference between both potentials follows for small $z$. In this area, the Yukawa potential models a more Woods-Saxon type potential, while the Riesz potential may be compared with an harmonic oscillator potential. But this behavior is restricted only to the lowest energy levels; for realistic calculations  the energy region near the Fermi-level is much more  relevant. In this region, both potential types  show a similar behavior for  $2.00 < \alpha < 2.50$.

Hence both potentials seem interesting candidates for generation of realistic singe particle energy levels.

For cylinder symmetric configurations the integral (\ref{convolAll}) cannot be solved analytically. Instead we switch to cylinder coordinates $\{\rho,z,\phi\}$. With the distance
\begin{equation}
\label{distcyl}
d_{\textrm{cyl}} = \sqrt{\rho^2 + {\rho}^{'2} -2 \rho {\rho}^{'}\cos(\phi^{'})+(z-z^{'})^2}
\end{equation}
we have to solve the integral
\begin{equation}
\label{convolAllcyl}
 V_{\textrm{type}}(\rho,z) =  C_{\textrm{type}} \int_{V}  d\rho^{'}\rho^{'}  dz^{'} d\phi^{'}  \rho(\rho^{'},z^{'}) w_{\textrm{type}}(d_{\textrm{cyl}})
\end{equation}
numerically.

In figure \ref{Rieszsph3} we have solved (\ref{convolAllcyl}) and compare the three different weights for the strong asymmetric  cluster decay 
\begin{equation}
^{232}\textrm{U} \rightarrow ^{208}\textrm{Pb} + ^{24}\textrm{Ne}
\end{equation}
The Riesz potential allows for a smooth transition between the Coulomb case  and the Yukawa limit by varying $\alpha$. Hence we obtain a direct geometric interpretation of the fractional parameter $\alpha$. 

Up to now, we discussed the properties of the single particle potential, which is the starting point for a calculation of the  microscopic part of the collective potential. 
 
On a macroscopic level the self energy of a given configuration contributes to the macroscopic part of the nuclear potential  as the Coulomb-  and surface or more sophisticated Yukawa energy term in a macroscopic energy formula, historically first used in  Weizs\"ackers famous liquid drop mass formula \cite{wei35}:
\begin{eqnarray}
\label{BWmass}
E_{\textrm{macro}} &=& a_v A + a_s A^{2/3} - a_c Z A^{-1/3} + \nonumber \\
&& a_{\textrm{sym}}(N-Z)A^{-1} + a_{\textrm{pair}} A^{-1/2}
\end{eqnarray}
as a function of the nucleon number $A =r_0 R_0^3$ containing a volume, surface, Coulomb,  symmetry and pairing term.
\begin{figure}[tbhp]
\begin{center}
\includegraphics[width=82mm]{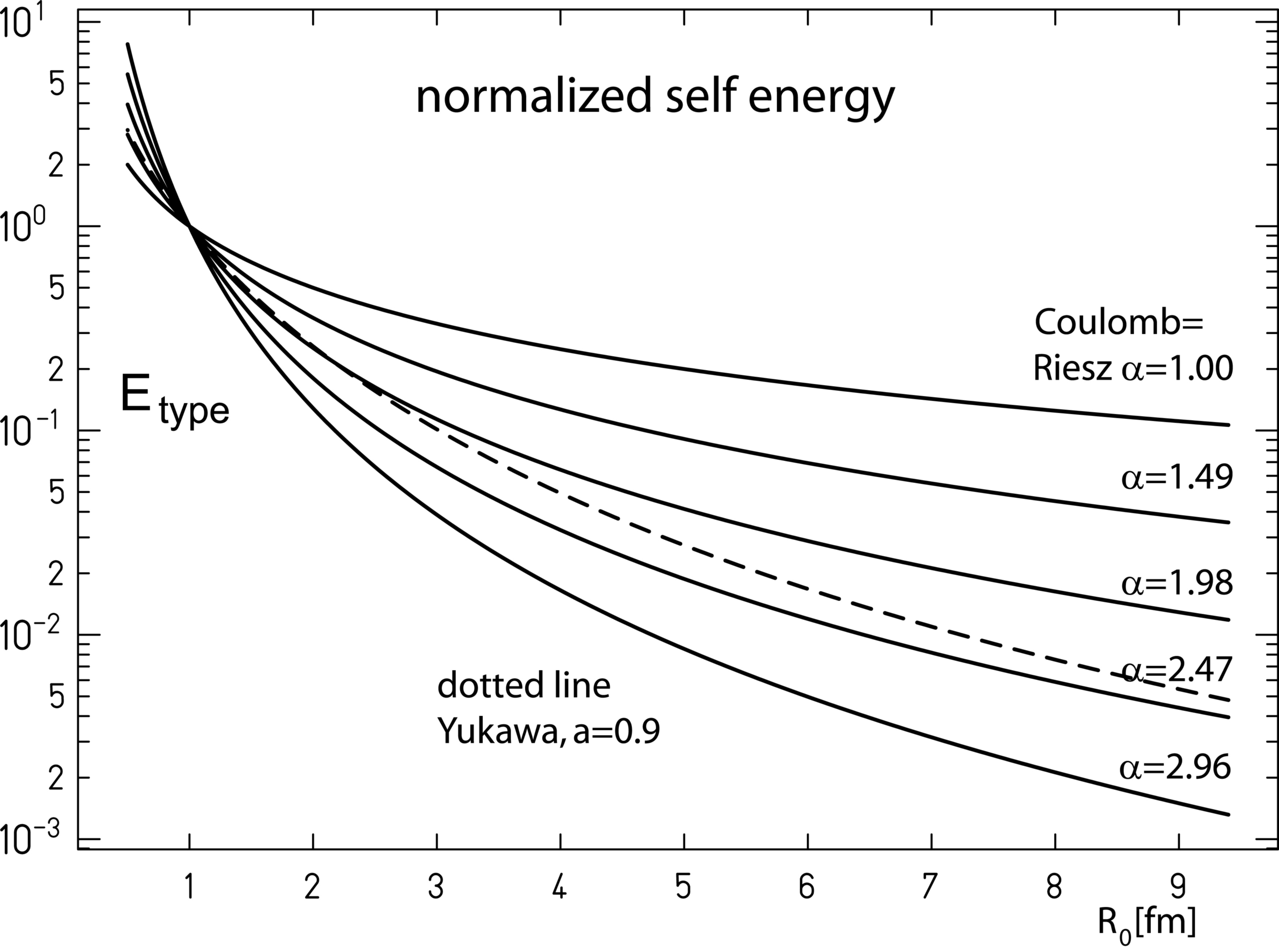}\\
\caption{
\label{selfE10}
{
For a spherical shape the self energy as a function of the sphere radius $R_0$[fm] is plotted for the Coulomb, the Riesz ($\alpha=1.49,1.98,2.47,2.96$ and the Yukawa ($a=0.9[fm]$ weight is plotted. To compare all different types, all energies are normalized to $E_{\textrm{type}}(R_0=1) = 1$. Depending on $R_0$, the Yukawa self energy lies is covered by the Riesz self energy within a range of $2.0 \leq \alpha 2.5$  values 
} }
\end{center}
\end{figure}

The self energy for a given charge type  $E_{\textrm{type}}$ is defined as the volume integral over the potential restricted on the volume of a given shape. 
\begin{equation}
\label{convolAllE1}
 E_{\textrm{type}}(\vec{r}) =  \frac{1}{2} C_{\textrm{type}} 
\int_{V}  d^3r \rho(\vec{r})
\int_{R^3}  d^3r^{'} \rho(\vec{r}^{'}) w_{\textrm{type}}(|\vec{r}-\vec{r}^{'}|)
\end{equation}
or
\begin{equation}
\label{convolAllE2}
 E_{\textrm{type}}(\vec{r}) =  \frac{1}{2} C_{\textrm{type}} 
\int_{V} \int_{V^{'}}  d^3r \,
d^3r^{'} \rho(\vec{r}) \rho(\vec{r}^{'}) w_{\textrm{type}}(|\vec{r}-\vec{r}^{'}|)
\end{equation}
For the 3 different weights (\ref{conweight1})-(\ref{conweight3}) we obtain for the simplest case of a sphere with radius $R_0$ , a unit charge ($Z e = 1$) and  with a thickness parameter  $a>0$:
\begin{eqnarray}
E_C      &=&      \frac{3}{5} \frac{a }{R_0}  \\
E_{RZ} &=&   \frac{9 \times 2^{2 - \alpha} a^\alpha }{(3-\alpha)(4-\alpha)(6-\alpha) }  \frac{1}{R_0^{\alpha}} ,  \quad 0\leq \alpha < 3   \\
E_{Y} &=&     \frac{3 a^3 - 3 a R_0^2 + 2 R_0^3 - 3 a e^{-2 R_0/a} (a + R_0)^2}{4 R_0^6} 
\end{eqnarray}

We obtain the important result that the Riesz self energy behaves like
\begin{eqnarray}
E_{RZ} &\sim& \frac{1}{R_0^{\alpha}} 
\end{eqnarray}
which scales with the nucleon number $A \sim R_0^3$ as
\begin{eqnarray}
E_{RZ} &\sim& A^{-\alpha/3} 
\end{eqnarray}
and therefore allows to model the influence of a screened Coulomb like charge contribution to the total energy.

In figure \ref{selfE10} we compare the $R_0$ dependence for the 3 different types of self energy. Depending on the size of the spherical nucleus e.g. $R_0(^{208}Pb)=7.25$ the behavior of the Yukawa self energy is covered by the Riesz self energy for $\alpha$ within the range $2\leq \alpha \leq 2.5$ and therefore the Riesz self energy covers the full range of relevant categories.

In addition it should be mentioned, that for the case $\alpha=3/2$ the Riesz self energy behaves like $A^{-1/2}$ which emulates the pairing term and for $\alpha=3 $  the Riesz self energy behaves like $A^{-1}$ which is equivalent to the behavior of the proton-neutron symmetry term in the Weizs\"acker mass formula. 

Therefore the macroscopic pairing-, symmetry- and Coulomb contributions to the total energy content of a nucleus may be treated from a generalized view as different realizations of the same Riesz potential and are all determined in the same way for a given shape.

As a consequence, there is a one to one correspondence between a given change in the shape geometry and the dynamic behaviour of these energy contributions. On the other hand,  the hitherto abstract fractional coefficient $\alpha$ now may be interpreted within the context of cluster physics as a smooth order parameter with a well defined physical meaning for distinct $\alpha$ values. 

The combination of concepts and methods developed in different branches of physics, here demonstrated for the case of fractional calculus and cluster physics, has always led to new insights and improvements. As an additional  step based on this concept, which marks a possible direction for future research,  we may emphasize the convolution aspect of the Riesz integral, which may be interpreted within the framework of  linear system theory,  leading to new insights in large amplitude collective motion. 

\section{Conclusion}

From all these presented results we may draw the conclusion, that the Riesz potential may be considered as a promising  alternative approach to folded potentials, which are widely used to describe nuclear dynamics within the framework of a collective shell model.  

Of course, these potentials are only an alternative starting point to calculate fragmentation potentials based on a fractional integral definition, but it is indeed remarkable,  that the Coulomb-, pairing- and  symmetry part of the macroscopic energy contribution may be considered as  specific realizations of the fractional Riesz integral with the fractional parameters $\alpha \in \{1, 3/2, 3\}$ .

\begin{acknowledgments}
We thank A. Friedrich  for useful discussions.
\end{acknowledgments}

%

\end{document}